\documentclass{article}

\usepackage[dvips]{graphicx}
\usepackage{amssymb}

\newenvironment{references}%
{
\footnotesize \frenchspacing

\renewcommand{\AA}{Astron.\ Astrophys.}

\newcommand{\ApJ}{Astrophys.\ J.}

\newcommand{\PASP}{P.A.S.P.}

\newcommand{\MNRAS}{MNRAS}
\section{{\rm REFERENCES}}
\sloppy \hyphenpenalty10000
\begin{list}{}{\leftmargin1cm\listparindent-1cm
\itemindent\listparindent\parsep0pt\itemsep0pt}}%
{\end{list}\vspace{2mm}}

\def\TYLDA{~}
\newlength{\DW}
\newcommand{\refitem}[5]{\item[]{#1} #2%
\def\REFARG{#3}\ifx\REFARG\TYLDA\else, {\it#3}\fi
\def\REFARG{#4}\ifx\REFARG\TYLDA\else, {\bf#4}\fi
\def\REFARG{#5}\ifx\REFARG\TYLDA\else, {#5}\fi.}

\begin{document}

\title{Hydrodynamical Models of Accretion Disks in SU~UMa Systems}
\author{Kacper Kornet, Micha\l\ R\'o\.zyczka\\
\normalsize{N. Copernicus Astronomical Center, 00-716 Warszawa, Bartycka
18} \\
\normalsize{\texttt{e-mail: kornet@camk.edu.pl, mnr@camk.edu.pl}}}
\maketitle

\begin{abstract}
We numerically test the mode-coupling model (Lubow 1991a) of tidal
instability in SU~UMa systems. So far, all numercial models confirming
it have been based on SPH codes and isothermal equation of
state. In our paper we present Eulerians models, using both isothermal
approximation and the full energy equation. We also investigate
influence of different ways of mass transfer. While isothermal models
behave similarly to SPH simulations, the behaviur of models
with full energy equation is quite different, and the mode-coupling
model is not confirmed in this case.
\end{abstract}

\section{Introduction}
SU UMa stars form a subclass of dwarf novae, which, in turn, are a
subclass of cataclysmic variables (CVs). Like all CVs, the SU UMa
stars are semidetached binary systems consisting of a white dwarf (the
primary) and a low-mass, main-sequence star (the secondary). The
secondary fills its Roche lobe, and a stream of gas flows from its
surface toward the primary through the inner Lagrangian point.
Because of excessive angular momentum, the stream is deflected from
its original direction, and an accretion disk is formed around the
primary. The disk may be subject to a thermal instability (Smak 1999),
resulting in episodes of enhanced accretion rate. To a distant
obsever, such an episode is visible as a temporary brightening of the
star, commonly referred to as an outburst.

As opposed to ordinary dwarf novae, the SU UMa stars exhibit a clearly
bimodal distribution of outbursts. Normal outbursts have an amplitude
of $\sim$ 3 mag, and last from one to four days. Superoutbursts are by
$\sim$ 1 mag stronger, and last for up to several weeks. The
recurrence time of normal outbursts (days to weeks) is not constant,
and it varies substantially form one system to another.  The
superoutbursts repeat more regularly, and their recurrence time is
much longer (months to years). In the extreme case of WZ Sge stars,
superoutbursts are nearly exclusively observed, with a recurrence time
of up to several tens of years.  In superoutbursts, the light curve of
a SU UMa system is modulated with a period a few per cent longer than
the orbital period. Those modulations are referred to as superhumps.
The superhump signal is known to originate from extended source(s) in
the outer disk (Warner 1995).

Superoutbursts are thought to be driven by a combination of thermal
and tidal instability. During normal outbursts, the disk grows in size
as it diffuses under the influence of increased viscosity. In systems
with mass ratios $q\lesssim0.25$, it eventually reaches up to the
location of the 3:1 resonance, at which the orbital frequency of the
disk gas is three times larger than the orbital frequency of the
binary (we define $q$ as the ratio $M_2/M_1$, where $M_1$ and $M_2$
are primary's and secondary's mass, respectively).  Subsequently, the
tidal instability sets in, the disk becomes eccentric, and, seen in
the inertial frame, it performs a slow, prograde precession. The tidal
influence of the secondary on (and the viscous dissipation in) the
outer disk, is largest when the bulk of the disk passes the
secondary. The superhump period is then the beat period between the
precession period and the orbital period of the binary (Osaki 1996).

Disk precession and superhump phenomenon have been subject to rather
intense theoretical investigations, largely based on numerical
simulations. According to Lubow (1991a,b), the eccentricity builds up
due to nonlinear interaction of waves, in which the m=3 component of
the tidal field is a key factor. Heemskerk (1994) performed
simulations using only that component, and he found that the disk
became eccentric, but it precessed retrogradely.  Moreover, with the
full tidal potential, the accretion disk was kept away from the
location of the resonance, and no significant eccentricity was
produced. Heemskerk's results are the only ones obtained with an
Eulerian (fluid) code. All remaining models presented in the
literature have been based on Lagrangian (particle) codes. A detailed
review of those calculations can be found in Murray (1998). While a
qualitative agreement with observations of superhump systems was
reached in several aspects, the models did not entirely agree with the
analytical theory of the tidal instability. In particular, the
measured eccentricity growth rates were much smaller than the
predicted ones. The models themselves were often based on an extremely
simple physical scenario (fully isothermal disk; gas from the
secondary uniformly "raining" onto a circular orbit within the disk).
In some of them the tidal instability was initiated by an arbitrary
increase of viscosity in the disk by a factor of 10. Finally, even in
the models which properly followed the stream of gas between the inner
Lagrangian point up to its collision with the edge of the disk, the
resolution in the collision region was too poor to resolve strong
shock waves responsible for the hot spot phenomenon.

In the present paper, we obtain Eulerian models of disks in SU Uma
systems in order to isolate the influence of various approximations on
the outcome of the simulations. The models can be directly compared to
Lagrangian models of Murray (1996, 1998). We perform both isothermal
simulations, and simulations in which the full energy equation with a
realistic cooling term is solved. We also compare "rainfall-type" mass
transfer models with those based on realistic modelling of the stream
from the secondary. The physical assumptions on which our models are
based are described in Sect.2 together with numerical methods emplyed
to solve the equations of hydrodynamics. The models are presented in
Sect. 3, and the results are discussed in Sect. 5.
 
\section{Physical assumptions and numerical methods}

We simulate the flow of gas in the orbital plane of a binary
consisting of two stars in circular orbits around the center of
mass. We use spherical coordinates cnd a corotating reference frame
centered on the primary, with the $z$-axis perpendicular to the
orbital plane. Assuming that the ratio $H/r$ of the disk is
constant, and the latitudinal velocity component is negligible, we can
write the continuity equation and the equations describing
conservation of radial and angular momentum in the following form:
\begin{eqnarray*}
\frac{\partial\rho}{\partial t} + \nabla\cdot(\rho \mathbf{\vec v}) &
= & 0 \\
\frac{\partial \rho v_r}{\partial t} + \nabla\cdot(\rho v_r \mathbf{\vec v}) &
= & \rho \frac{v_\phi^2}{r} - \frac{\partial p}{\partial r} + \rho f_r
+ F^{visc}_r\\
\frac{\partial j}{\partial t} + \nabla\cdot(j \mathbf{\vec v}) &
=& -\frac{\partial p}{\partial \phi} + r \rho f_\phi + r F^{visc}_\phi
\end{eqnarray*}
where for any variable $a$
\[
\nabla\cdot(a\mathbf{\vec v}) = \frac{1}{r^2} \frac{\partial r^2 a
  v_r}{\partial r} + \frac{1}{r} \frac{\partial a v_\phi}{\partial
  \phi}
\]
In these equations $j=r\! \rho v_\phi$ is the angular momentum density
measured in corotating frame, and $\mathbf{\vec f}$ is the external
force (gravitational and inertial) acting on unit volume. Viscous
forces $F^{visc}$ are given by standard formulae (see eg. Landau and
Lifshitz 1982). We assumed that kinematic viscosity coefficient and
bulk to shear coefficients ratio are constant troughout the disk.

The models are either isothermal or radiative. In the first case we
use an isothermal euation of state
\[
p = c_s^2 \rho,
\]
where $c_s$ is the isothermal sound speed, assumed to be constant in
space and time.  In the second case the equation of state of an ideal
gas with the ratio of specific heats $\gamma$ equal to $5/3$ is
used, and the energy equation
\begin{equation}
\frac{\partial E}{\partial t} + \nabla\cdot(E\mathbf{\vec v}) = 
                -p \nabla\cdot\mathbf{\vec v} +Q^{visc} - Q^{rad}
\label{ren}
\end{equation}
is additionally solved, with $E$ standing for the internal energy density,
$Q^{visc}$ - for heat generated by viscous forces and $Q^{rad}$ - for
heat radiated away. Cooling processes are described in the same way as
in R\'o\.zyczla and Spruit (1993).

The equations are solved with the help of an explicit Eulerian code
described by R\'o\.zyczka (1985) and R\'o\.zyczka and Spruit
(1993). The inner edge of the grid is located at $r_{in}=0.1 d$ from
the primary (where $d$ is the distance between the components of the
system), and the outer one ($R_{out}$) - at the inner Lagrangian point
$L_1$. For all models the same mass ratio $q=3/17$ is assumed,
resulting in $r_{L_1}=0.736 d$. The spacing of the grid is logarithmic
in $r$ and uniform in the azimuthal coordinate $\phi$, and there are
$50$ and $60$ grid points in $r$ and $\phi$, respectively.

The gas can freely flow into the computational domain or out of it
through the outer boundary of the grid. At the inner grid boundary
only free outflow is allowed for, and no inflow can occur. Due to the
explicit character of the numerical scheme, the length of the time
step is limited by the Courant condition. In the present simulations
the Courant factor is $0.3$.

\section{Initial setup and results of simulations}

In the present paper we repeat and refine two simulations performed by
Murray (1996), also referred to as models $4$ and $5$ in a later
discussion by Murray (1998). As mentioned in the Introduction, Murray
uses a Lagrangian SPH code, whwreas we work with an Eulerian, grid
based one. Additionally, we employ a more realistic physical scenario,
taking into account the effects of cooling, and using an ideal gas
equation of state instead of an isothermal one. Following Murray we
adopt $q=3/17$, a kinematic viscosity coefficient of $2.5 \times
10^{-4} d^2 \Omega_\ast$ (where $\Omega_\ast$ is the orbital frequency
of the binary), and a sound velocity of $0.02 \ d \ \Omega_\ast^{-1}$
for the isothermal cases.

As an initial condition we chose isothermal, Keplerian disk
with density varying as $r^{-1.5}$. The disk extends from the inner
boundary of the grid up $r=0.35d$. The rest of the grid is originally
filled with a rarefied ambient medium at Keplerian rotation around the
primary. The density of the ambient medium is 1\% 
of the disk density
at $r=0.35$, and its pressure matches that of the disk.

The original two models of Murray differ only in details of mass
transfer. In the first case the gas from the secondary enters the
computational domain as a narrow stream originating at $L_1$; in the
second case the gas "rains" uniformly onto the orbit corresponding to
the circularization radius. In the following, they are referred to as
{\it stream} and {\it rain} models, respectively (note that, by the
definition of the circularization radius, if the mass transfer rate is
the same in both cases then the rate of angular momentum transfer is
also the same). Below we describe four models: radiative stream,
radiative rain, isothermal stream, and isothermal rain. Hereafter,
they are referred to as Models (or Cases) 1 - 4, correspondingly.

All models were followed for 100 orbits of the binary. In order to
check if the balance between the matter added to the disk and accreted
onto the white dwarf is established in the course of evolution, we
monitored the disk mass as a function of time (Fig.1). For models 1, 2
and 3 the mass of the disk grows rapidly in the beginning, and then it
reaches an equilibrium value. That value is almost two times greater
in Case 4 (isothermal rain) than in Case 2 (radiative rain). In Case 3,
after the phase of initial growth, the mass of the disk begins to
oscillate with a period of $\sim45P$ (where $P$ is the orbital period
of the binary).
\begin{figure}
\begin{center}
  \includegraphics[angle=270,width=0.9\textwidth]{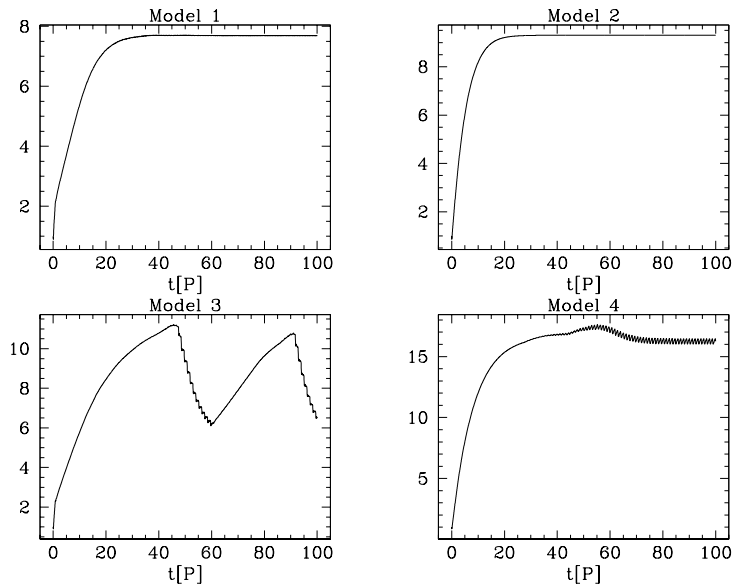}
\caption{\small The mass of the gas contained in the grid as a function of
  time. The plots are normalized to the initial mass of the disk.}
\end{center}
\label{masy}
\end{figure}

\begin{figure}
  \begin{center}
  \includegraphics[angle=270,width=0.9\textwidth]{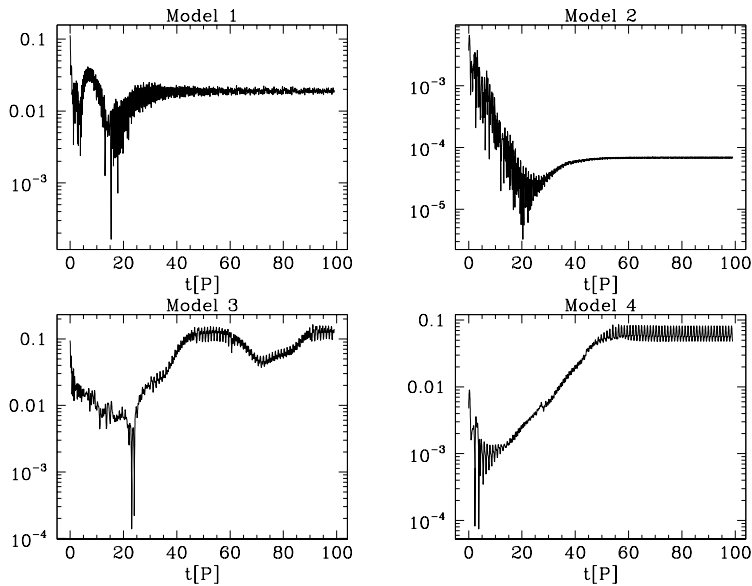}
  \caption{\small $S_{1,0}/S_{0,0}$ as a function of time.}
  \label{mod_10} \end{center}
\end{figure}

\begin{figure}[t]
  \begin{center}
  \includegraphics[angle=270,width=0.9\textwidth]{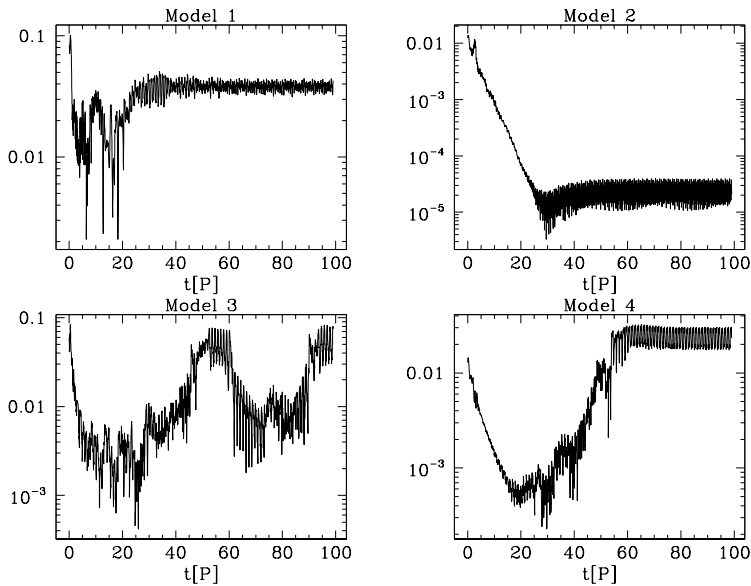}
  \caption{\small $S_{2,3}/S_{0,0}$ as a function of time.}
  \label{mod_23} \end{center}
\end{figure}

The next step was to check, whether the disks in our models became
eccentric. For natural numbers $k,l$ we can calculate
\[
S_{k,l}=\left| \int_{r_{in}}^{r_{out}} \rho_{k,l} \, r^2 dr \right|,
\]
where $\rho_{k,l}$ are defined by equation
\[
\rho=\sum_{k=0}^{\infty} \sum_{l=0}^{\infty} \rho_{k,l}
\exp [i(k \theta - l \Omega_\ast t)]
\]
and the term $r^2$ results from the assumption of constant 
ratio $H/r$ (in practice, the lower integration limit can be equal to
$r_{in}$, whereas the upper one has to be placed at $r=0.6$ in order
to avoid highly asymmetric contribution from the stream). If the mass
of the disk is constant, then an appropriate measure of the
eccentricity is $S_{1,0}$. If, however, the mass is varying in the
course of simulation, $S'_{1,0}$ should be used instead, where
$S'_{1,0} \stackrel{def}{=} S_{1,0}/S_{0,0}$ (note that $S_{0,0}$ is
proportional to the mass of gas in the integration
domain). Fig. \ref{mod_10} shows $S'_{1,0}$ as a function of time.
It can be immediately seen that $S'_{1,0}$ is much greater in
isothermal models than in radiative ones. In other words, isothermal
disks more willingly become elliptic. Secondly, only Model 4
(i.e. isothermal rain) exhibits a phase of exponential growth of
$S'_{1,0}$ which was predicted by Lubow's theory. Even in that case,
however, another prediction of the theory is not met, namely, the
behavior of the (2,3) mode.  The strength of that mode should be
proportional to the time-derivative of the strength of (1,0) mode,
whereas in our simulations it is roughly proportional to the strength
of (1,0) mode  (see Fig.  \ref{mod_23}).

\begin{figure}
  \begin{center}
  \includegraphics[angle=270,width=0.8\textwidth]{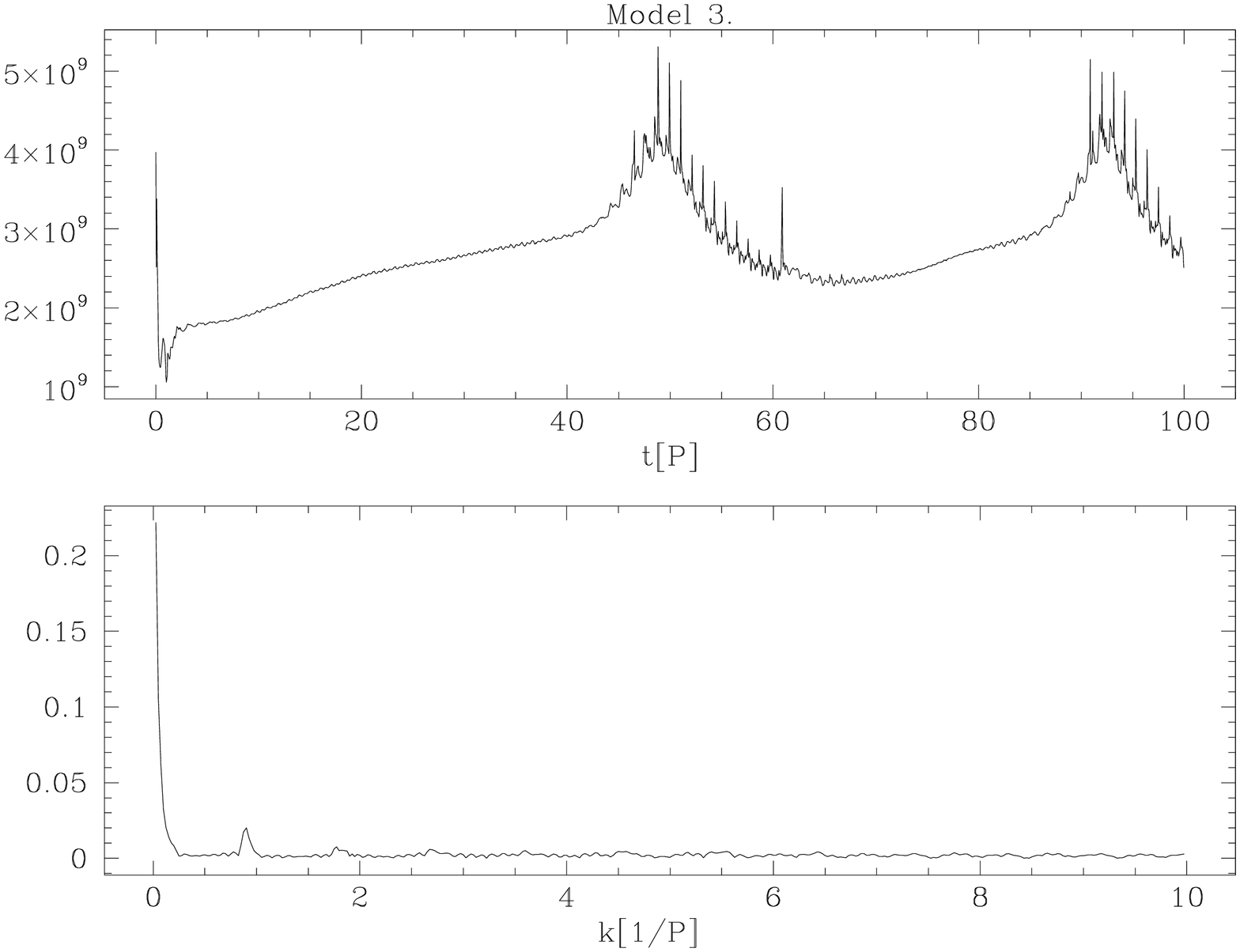}
  \caption{\small Viscous dissipation rate (program units) in Model 3 
  as a function of time (top), and its power spectrum in time range of
  $60-100P$ (bottom).  dissipation rate in this time range.
  \label{enlep_izo}} \end{center}
\end{figure}

\begin{figure}
  \begin{center}
  \includegraphics[angle=270,width=0.8\textwidth]{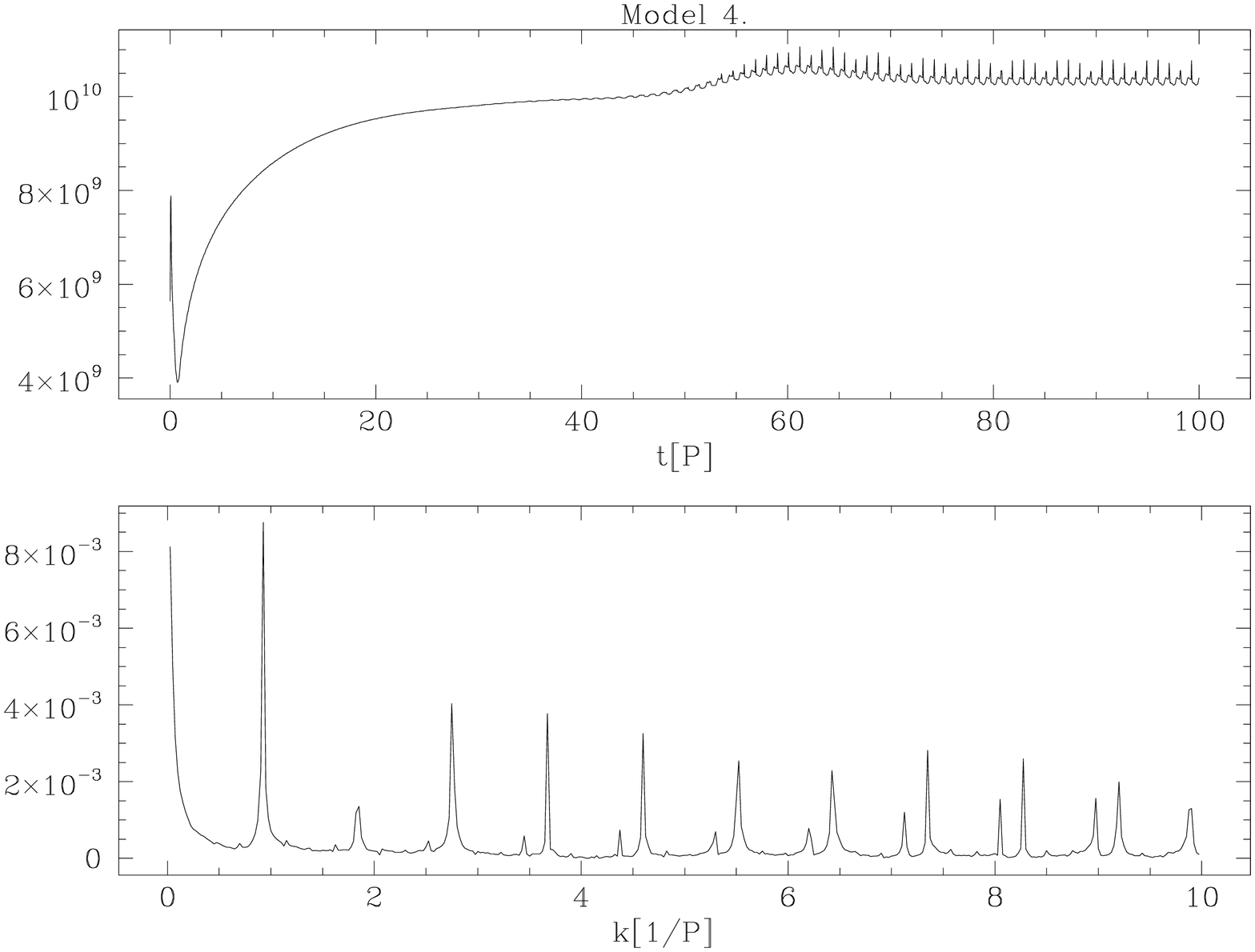}
  \caption{\small Viscous dissipation rate (program units) in Model 4 
  as a function of time (top), and its power spectrum in time range of
  $60-100P$ (bottom). \label{enlep_izokepl}} \end{center}
\end{figure}

Following Murray (1996, 1998), we calculated the rate of viscous
dissipation in the disk as a function of time. The results of Foureir
analysis of that function for Models 3 and 4 in range of $60-100P$ are
shown in Fig. \ref{enlep_izo} and \ref{enlep_izokepl}.
In Case 4 (isothermal rain) the rate of viscous dissipation reaches a
stationary value, whereas in Case 3 (isothermal stream) it exhibits
large amplitude oscillations (by a factor of 2) with a period of $\sim
45P$. Note that the maxima of the dissipation rate coincide with the
maxima of the the disk mass. In both models much more rapid
oscillations of the dissipation rate are also seen, whose dominant
frequency is a little lower than $1P^{-1}$ (all remaining peaks in
power spectra can be interpreted as its higher harmonics). The
corressponding periods are equal to $1.11P$ in Model 3., and $1.08P$
in Model 4. In Model 3 the high-frequency oscillations are visible
only within dissipation rate maxima (on the ascending branch shortly
before peak, and on the entire descending branch). In Model 4 they are
excited when the mass of the disk approaches its maximum, and they
persist till the end of simulation. A detailed examination of i
high-frequency oscillation reveals their similarity to those found by
Murray (1998) in his Models 4 and 5 (see his Fig. 6). Because they are
visible both in model with and without stream, those oscillations
cannot be excited by the interaction of the stream with the advancing
and retreating edge of the disc, and
they most probably must be related to the tidal influence of the
secondary. We found no similar oscillations in our radiative models.

\begin{figure}
  \begin{center} \includegraphics[width=0.9\textwidth]{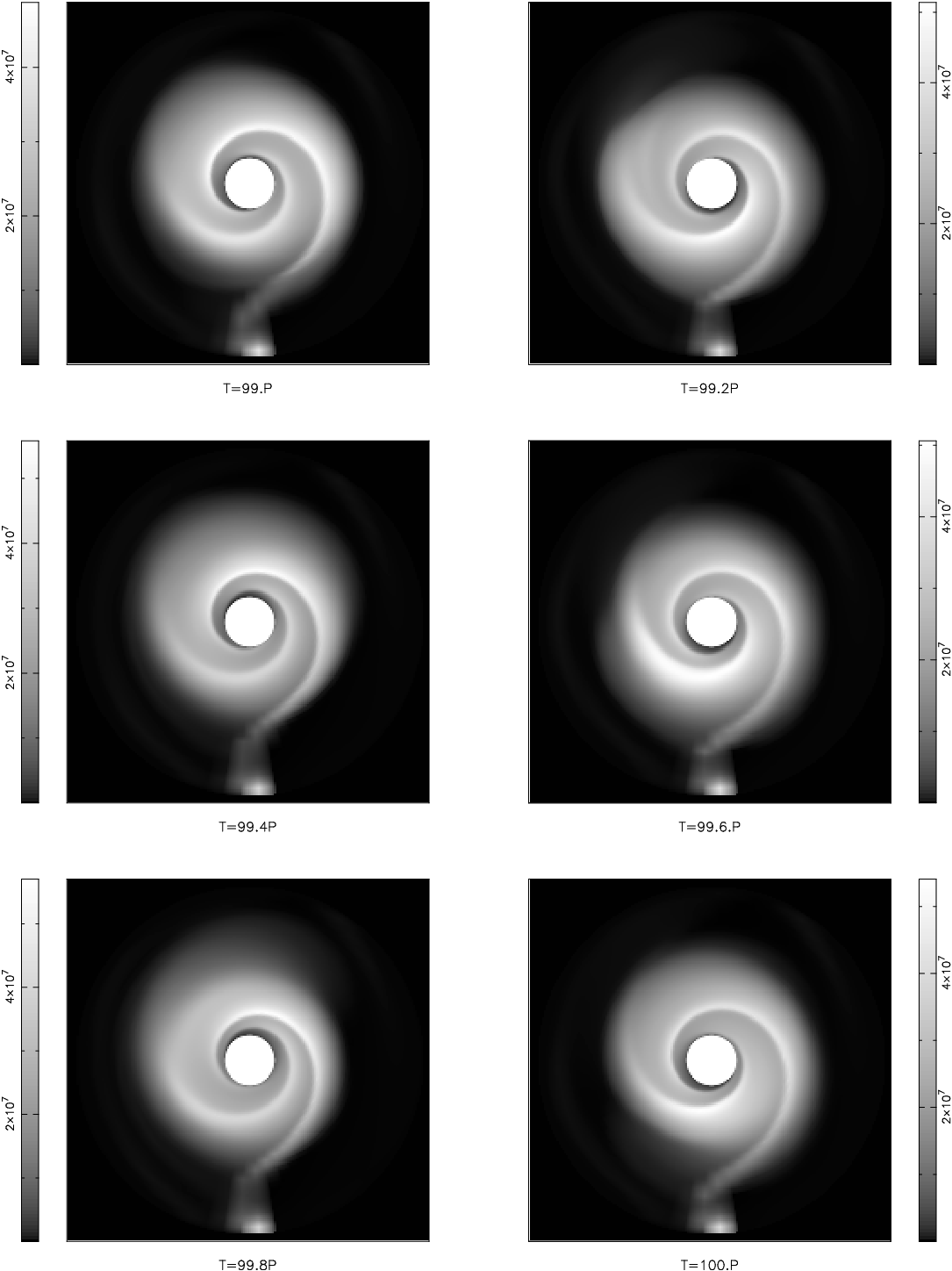}
  \caption{\small Density maps for Model 1. The snapshots are taken at
  $t =$ $99.P$, $99.2P$, $99.4P$, $99.6P$, $99.8P$ and $100P$. The
  white dwarf and the $L_1$ point are at the center, and in the middle
  of the lower edge of each frame.}  \label{ch99} \end{center}
\end{figure}

\begin{figure}
  \begin{center} \includegraphics[width=0.9\textwidth]{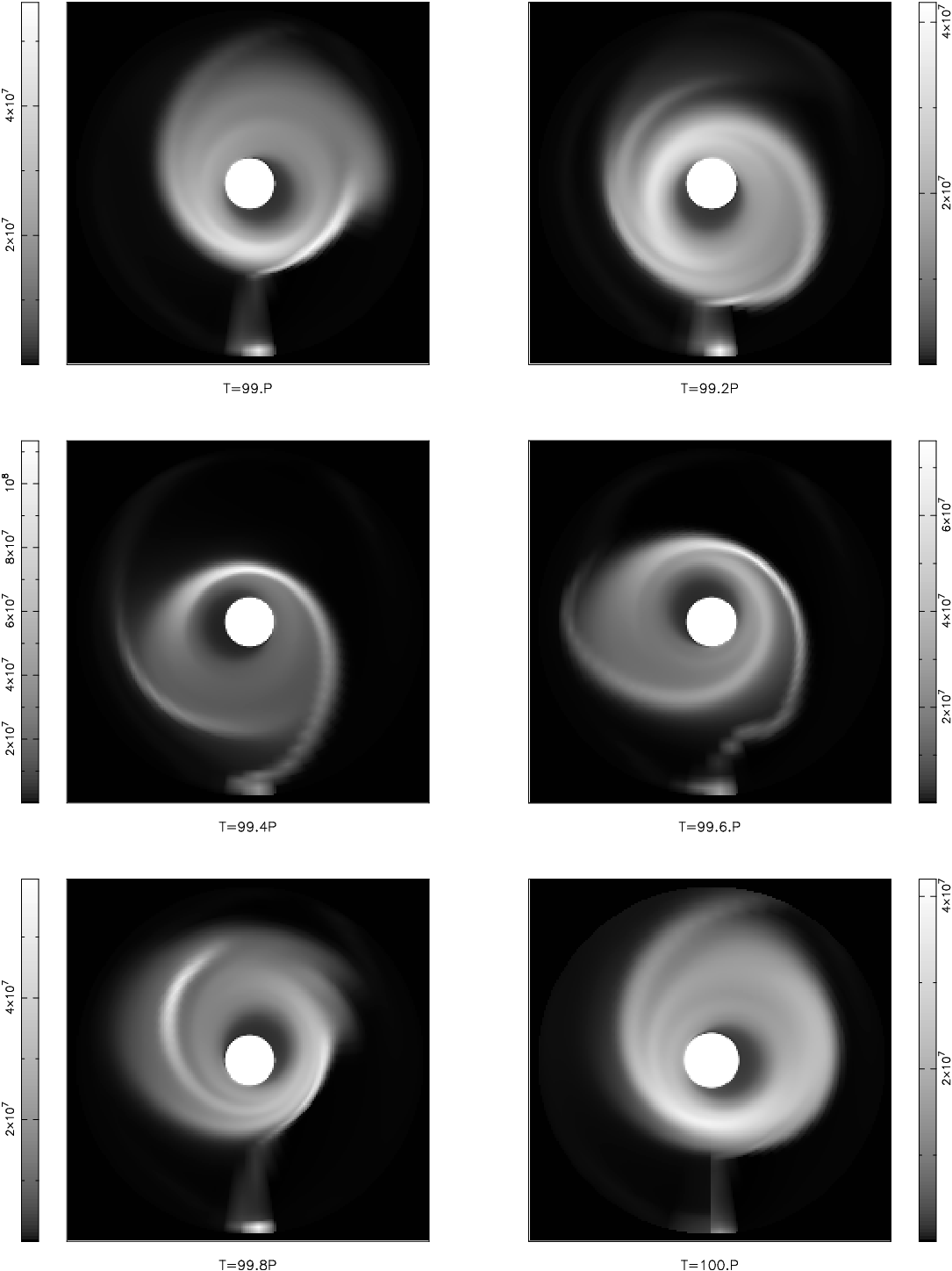}
  \caption{\small Density maps for Model 3. The snapshots are taken at
  $t =$ $99.P$, $99.2P$, $99.4P$, $99.6P$, $99.8P$ and $100P$. The
  white dwarf and the $L_1$ point are at the center, and in the middle
  of the lower edge of each frame.}  \label{izo99} \end{center}
\end{figure}

In Fig. \ref{ch99} and \ref{izo99} we present sequences of disk
density maps for Models 1 and 3 (radiative stream and isothermal stream),
covering the time-span from $99P$ to $100P$ (at that time the rapid
oscillations of viscous dissipation rate are fully developed in Model
3).
In the radiative case the disk is almost circular. However, in the
isothermal case the disk has a clearly nonaxisymmetric shape, which,
at least in some frames, does not significantly differ from the
elliptical one. The ellipse is precessing with a period a little
longer than $1P$. In one precession period the maximum viscous
dissipation occurs when the major axis of the ellipse is perpendicular
to the line connecting the components of the binary ($t=99.65P$ in
Fig. \ref{izo99}) (the same results were obtained by Murray). In both cases a
nonaxisymmetric feature in the form of two spiral arms is also seen,
which in Case 1 remains stationary in the corotating frame. We
interpret it as the $(2,2)$ mode, excited and maintained by the tidal
interaction of the secondary.

\begin{figure}
  \begin{center} \includegraphics[width=0.9\textwidth]{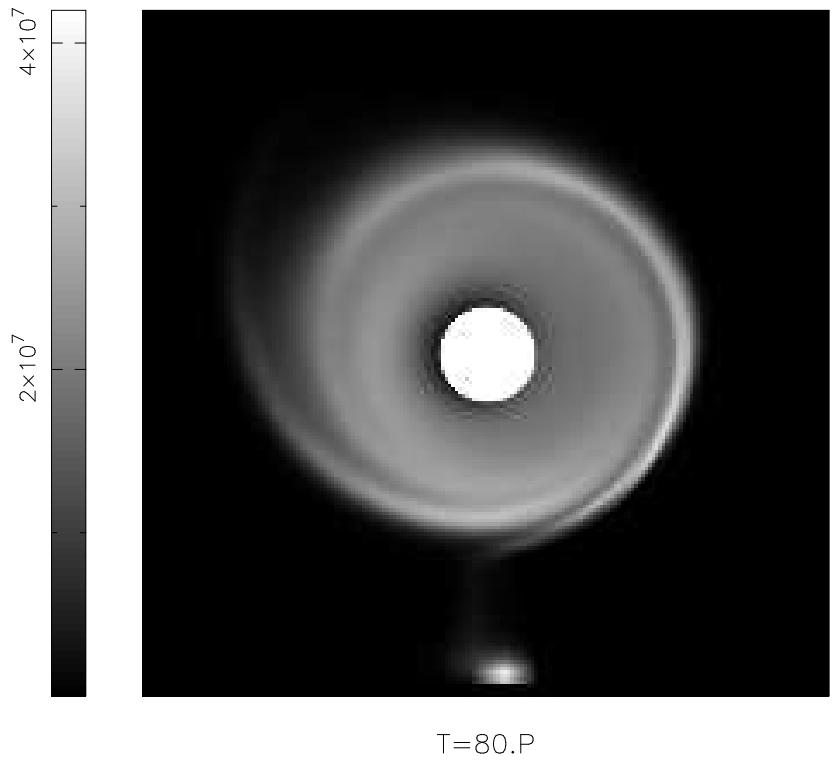}
  \caption{\small Density map for model 3. at the time $t=80P$.  The
  white dwarf and the $L_1$ point are at the center, and in the middle
  of the lower edge of the frame.}
\label{izo80} \end{center}
\end{figure}

Fig. \ref{izo80} shows the density map of Model 3 at $T=80P$
(i.e. between the large-scale maxima of the dissipation rate, when no
short-period oscillations are visible).
At that moment the disk is nearly axisymmetric, and only the (2,2)
mode has a significant amplitude. Thus, we conclude that the overall
behaviour of Model 3 is in agreement with the tidal instability model
of superhumps.

\section{Discussion}
In the present paper we reported a series of 4 simulations of disk in
Cataclysmic Variables of SU UMa type. The least realistic model was
obtained with an isothermal equation of state, and a "rainfall-type"
approximation for mass transfer from the secondary. The most realistic
simulation involved a full energy equation, and a stream of matter
originating at the inner Lagrangian point.

We found that only the isothermal disks exhibit a clear tendency
toward elliptic distortions accompanied by precession. Within the
framework of Lubow's (1992) theory, one could try to explain this
result by assuming, that the strength of the tidally excited
(2,2) mode is significantly larger in radiative models than in the
isothermal ones. This is because, according to the theory, the (2,2)
oscillations tend to keep the disk gas away from the 3:1 resonance
responsible for the growth of the tidal instability. Such an
assumption, however, is not confirmed by the analysis of our results:
the (2,2) mode appears to be equally strong in all models.

The phase of an exponential growth of the (1,0) mode, foreseen by the
theory, was found only in the least realistic model. However, contrary
to theoretical predictions, even in that case the time derivative of
the strength of (1,0) mode was not proportional to the strength of the
(2,3) mode.

Oscillations with a period slightly longer than $P$, which may be
tentatively associated with superhumps, were observed in isothermal
models only. The period of oscillations found in radiative models was
about 3 times shorter.

The main results of this work may be summarized in three points:

1. As foressen by the tidal instability theory, isothermal develop an
appreciable eccentricity, and begin to precess. The precession period
is the same as the period of rapid fluctuations in viscous dissipation
rate, and it is slightly longer than the orbital period of the binary.
The behaviour of the (1,0) mode, however, is not consistent with the
theory.

2. In radiative models an elliptic, precessing disk does not
develop. The dominant oscillations in the viscous dissipation curve
have periods of $\sim 1/3P$.

3. In all models the two-armed (2,2) mode, excited and maintained by
tidal forces of the secondary, is very clearly seen.

Our conclusion is that the mechanism of superhump phenomenon is not
yet entirely understood, and further research on this cubject is
desirable. R\'o\.zyczka \& Spruit (1993) simulated a viscosity-free
disk with radiative losses, in which angular momentum was transported
by spiral shocks. They found an eruptive instability, qualitatively
similar to outbursts of Dwarf Novae. During the eruption, the radius
of the disk increased substantially. We suggest that low-viscosity
disks prone to this type of instability may develop eccentricity and
exhibit superhump-like oscillations during outbursts. This issue is
presently under investigation.

{\bf Acknowledgments} This research was supported by the Committee for 
Scientific Research through the grant 2.P03D.004.13.

\end{document}